\documentclass[10pt]{article}

\usepackage{fullpage}
\usepackage{amsmath,amsfonts,amssymb}
\usepackage{graphicx}
\usepackage{setspace}
\usepackage{xcolor,colortbl}
\usepackage{tocloft}
\usepackage{hhline}
\usepackage{color}
\usepackage{bm}  
\usepackage{hyperref}
 \usepackage{stackrel}
 \usepackage{upgreek}
 \usepackage{csquotes}
 \usepackage[sans]{dsfont} 
 \usepackage{cite}
\usepackage{lineno}
 \DeclareGraphicsExtensions{.pdf,.eps,.png,.jpg,.mps}

\def\=#1{\underline{\underline #1}}
\def\##1{{\bf #1}}

\def\red{\textcolor{black}}

\def\le{\left(}
\def\ri{\right)}

\def\lec{\left\{}
\def\ric{\right\}}

\def\eps{\varepsilon}

\def\lambdao{\lambda_{\scriptscriptstyle 0}}

\def\ko{k_{\scriptscriptstyle 0}}

\def\epsmet{\eps_{\rm met}}
\def\epsdiel{\eps_{\rm diel}}

\def\epsoref{\=\eps_{\rm ref}^{\rm o}}
\def\epsref{\=\eps_{\rm ref}}
\def\epsa{\eps_{\rm a}}
\def\epsb{\eps_{\rm b}}
\def\epsc{\eps_{\rm c}}

\def\nmet{n_{\rm met}}
\def\ndiel{n_{\rm diel}}
\def\nprism{n_{\rm prism}}
\def\ninf{n_{\rm inf}}
\def\nwater{n_{\rm water}}

\def\Lmet{L_{\rm met}}
\def\Ldiel{L_{\rm diel}}

\def\ux{\hat{\#u}_{\rm x}}
\def\uy{\hat{\#u}_{\rm y}}
\def\uz{\hat{\#u}_{\rm z}}

\def\.{\mbox{ \tiny{$^\bullet$} }}

\def\Einc{{\#E}_{\rm inc}({\#r})}
\def\Erefl{{\#E}_{\rm ref}({\#r})}

\def\cpsi{\cos\psi}
\def\spsi{\sin\psi}
\def\ctheta{\cos\theta}
\def\stheta{\sin\theta}

\def\thetaSPP{\theta_{\rm SPP}}
\def\as{a_s}
\def\ap{a_p}
\def\rs{r_s}
\def\rp{r_p}

\cftpagenumbersoff{figure}
\cftpagenumbersoff{table}

\begin{document} 

\begin{center}
\textbf{Artificial neural network to estimate the refractive index of a liquid infiltrating a chiral sculptured thin film}\\

{Patrick D. McAtee,$^{a,\ast}$
Satish T.S. Bukkapatnam,$^b$
 and Akhlesh Lakhtakia$^a$}\\

$^a$The Pennsylvania State University,  Department of Engineering Science and Mechanics, University Park, PA 16802, USA\\
$^b$Texas A\&M University, Department of Industrial and Systems Engineering, College Station, TX 77843, USA
\end{center}

\begin{abstract}
We theoretically expanded the capabilities of optical sensing based on surface plasmon resonance  in a
prism-coupled configuration by incorporating artificial neural networks (ANNs). We used calculations modeling
 the situation in which
an index-matched substrate with   a metal thin film and a porous chiral sculptured thin film (CSTF) deposited
successively on it is affixed
 to the base of a triangular
prism. When a fluid is brought in contact with the exposed
face of the CSTF, the latter is infiltrated. As a result of infiltration, the traversal of light entering one slanted face of the prism
and exiting the other slanted face of the prism is affected.  We trained
two ANNs with differing structures using reflectance data generated from
simulations to predict the refractive index  of the infiltrant fluid. The best predictions were a result of training the ANN with simpler structure. With realistic simulated-noise, the performance of this ANN is robust.
\end{abstract}


 \section{Introduction}
\label{sect:Intro}  

The ability to accurately  detect small concentrations of chemicals and biochemicals, whether toxic or benign,  is highly prized in chemical, pharmaceutical, medical, environmental, and food industries \cite{McDonagh2008,Turner_2013}. Introduction of pathogens and toxins into the human body can arise from intentional contamination of essential infrastructure \cite{Lim_2005} as well as from unforeseen consequences of their otherwise necessary applications \cite{Verma_2014}. Even chemicals traditionally thought of as non-toxic in their bulk form,  such as gold, could have harmful effects when ingested as nanoparticles \cite{Orlando_2016,Fratoddi_2015,Alkilany_2010}. Also, the concentrations of various chemicals in solutions and dispersions need to be determined in research as well as industrial laboratories \cite{McDonagh2008,Homola2008,Malekzad2018}.

Sensors of chemicals and biosensors are designed to operate on the basis of several different phenomena, including electrochemical \cite{Stetter_2008,Cosnier_2015}, optical \cite{Abdulhalim_2008,Taliercio_2017,Arjmand_2017}, piezoelectric \cite{Skladal_2016}, gravimetric \cite{Ramos_2017}, and pyroelectric \cite{Davidson_2017}. Our focus here lies on optical  sensors,
of which several types exist \cite{Rasooly,Zourob}.

One commonly used optical-sensing technique relies on surface plasmon resonance (SPR) which can occur when light interacts with free electrons at a metal/dielectric interface \cite{Homola2008}. As a result, a surface-plasmon-polariton (SPP) wave is excited.  Changing the \red{relative} permittivity of the dielectric material will change the characteristics of the SPP wave, thus allowing for sensing \cite{PMLbook}. Let us consider an SPP wave propagating along the $x$ axis guided by
the metal/dielectric interface  $z = 0$ and suppose that the metal and the partnering dielectric material are isotropic and homogenous. The metal of relative permittivity $\epsmet$ fills the half-space $z < 0$ and the partnering dielectric material of relative permittivity $\epsdiel$ fills the half-space $z > 0$. The complex-valued wavenumber of the SPP wave guided by the interface $z=0$ is
given by  
 \begin{equation}
q = \ko \sqrt{\epsdiel \epsmet/(\epsdiel + \epsmet)}\,,
\end{equation} 
where $\ko$ is the free-space wavenumber. This SPP wave is $p$ polarized \cite{Simon}.

In order to evoke SPR in practice, many geometric configurations have been devised to couple  incident light to the SPP wave guided by the metal/delectric interface. The most commonly implemented configuration is a prism-coupled configuration called the Turbadar--Kreschmann--Raether (TKR) configuration \cite{Turbadar_1959, KR_1968}, wherein
a metal film of thickness $\Lmet$ and a dielectric film 
of thickness $\Ldiel$ are successively deposited onto one face of a substrate, thus establishing a \textit{sensor chip} \cite{Swiontek_2016}. The substrate
has the same refractive index as a prism of refractive index $\nprism$ which exceeds $\ndiel = \sqrt{\epsdiel}$,
our assumption being that both $\nprism$  and  $\ndiel$ are real and positive.
Typically, the prism has a cross-section of a $45^\circ$--$90^\circ$--$45^\circ$ triangle. The second face of the substrate is affixed to the hypotenuse of the prism using an index-matching fluid.
A monochromatic $p$-polarized plane wave of
free-space wavelength $\lambdao$ and intensity $I_0$ is incident onto one slanted face of the
prism at an angle $\phi$ with respect to the normal to that face. The refracted plane wave is incident on the substrate/metal interface
(effectively, the prism/metal interface) at an angle $\theta$  with respect to the normal to the interface; in principle,
$\theta\in [0^\circ , 90^\circ)$. The plane wave is reflected and exits the other slanted face of
the prism. The intensity $I_r$ of the exiting plane wave  is measured as a function of $\theta $   by a photodetector, and thereby the
reflectance  $R = I_r/I_0$
 is deduced as a function of $\theta$.
A sharp dip
in the graph of $R$ vs. $\theta$ indicates the excitation of a SPP wave at a specific angle denoted by $\thetaSPP$,
provided that $\sin\thetaSPP>\ndiel/\nprism$.
This angle is characteristic of $\epsdiel$ and is related to the SPP wavenumber as follows:
\begin{equation}
\text{Re}(q) \simeq \ko \nprism \sin \thetaSPP\,.
\end{equation}
The angle $\thetaSPP$  changes with $\ndiel$. This is the principle of SPR-based sensing,
with the partnering dielectric material being the material whose refractive index is the quantity sensed.
 
 Only one dip indicating SPR appears at a fixed free-space wavelength $\lambdao=2\pi/\ko$, because the chosen metal/dielectric interface can guide just one SPP wave. Only one SPP wave can be excited at a fixed $\lambdao$ even if the partnering dielectric material is anisotropic \cite{Sprokel_1981_1,Sprokel_1981_2}. 
 
 Theory \cite{PMLbook} and subsequent experiments \cite{Devender,Lakhtakia_2009,Gilani_2010},
however, have confirmed that  a periodically nonhomogeneous dielectric material, whether isotropic  or anisotropic, partnering a metal in the TKR configuration can support the existence of multiple SPP-wave modes at a fixed $ \lambdao$. The different SPP-wave modes have their peak field within the partnering dielectric material at different distances from the interface  \cite{PMLbook}. If the periodically nonhomogeneous dielectric material is porous and is infiltrated by a fluid of refractive index $\ninf$, theory shows \cite{Mackay_2012} and experiment has confirmed \cite{Swiontek_2013,Swiontek_2016} that the locations of all SPP-wave modes    in the graph of $R$ vs. $\theta$ shift with $\ninf$, thereby enabling  optical-sensing applications.

In addition to SPP-wave modes, waveguide modes \cite{Khaleque,Marcuse}
may also manifest in the graph of $R$ vs. $\theta$. These waveguide modes 
differ from SPP-wave modes in that they can be bound to more than one interface, and therefore depend on the thickness of the partnering dielectric material \cite{Liu_2015}.

There could even exist signatures of undiscovered phenomena in the graph of $R$ vs. $\theta$. Therefore, when deducing what $\ninf$ generated a specific  graph of $R$ vs. $\theta$, it is best to make use of all the features of the graph. This calls for the use of artificial neural networks (ANNs), which do not require understanding of the various underlying phenomena to discern a complicated quantitative relationship between them \cite{Goodfellow_2016}.

Currently, ANNs and other machine-learning algorithms are being applied to a multitude of tasks in engineering and medicine \cite{Bukk_1997, Bukk_2000, Bukk_1999, Bukk_1995, Braun_2004,Chak_2008,Saetchnikov_2012,Rogers_2013,Mutter_2017,Maleki_2017}, 
\red{including inverse optical design \cite{Ma_2018,Liu_2018}.}
\red{Although ANNs and machine learning have been applied to SPR biosensors  \cite{Nezhad_2010,Ma_2017,Nezhad_2015,Yu_2019,Bahrami_2013}}, their use in scenarios to exploit the excitation of multiple guided-wave modes (including multiple SPP wave-modes and waveguide modes) as well as of other polarization-dependent features in the graph of $R$ vs. $\theta$ provides a novel avenue for optical sensing.

The plan of this paper is as follows. Section~\ref{sect:tp} provides brief introductions to: (i) the calculation of reflectances when a porous chiral sculptured thin film (CSTF) \cite{Mackay_2012,Swiontek_2013,Lakh2006a,Lakh2006b} is used as the partnering dielectric material
in the TKR configuration, and (ii) our application of ANNs to that optical-sensing scenario \cite{confpaper}.
 In Sec.~\ref{sect:Exp}, we give the parameters of two different ANNs devised by us and  the data used to train and test each ANN. Section \ref{sect:Results} details the performance of each ANN, and Sec.~\ref{sect:Conclusion} contains a discussion
of  the implications of the numerical results.

\section{Theoretical Preliminaries}\label{sect:tp}

\subsection{Optical Sensing}
\label{sect:os}

\begin{figure}
\centering
\includegraphics[scale=0.7]{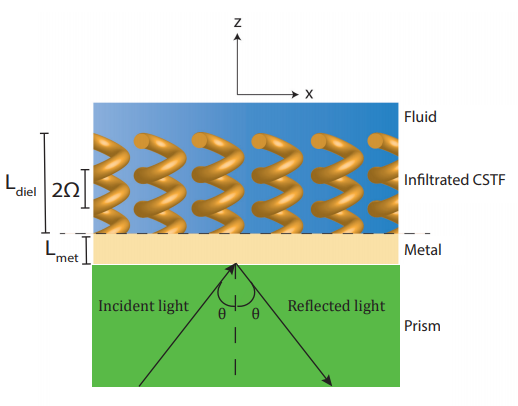}
\caption{Schematic of the TKR configuration used to calculate $R$ vs. $\theta$ when the partnering dielectric material is a CSTF infiltrated by a fluid of refractive index $\ninf$. All layers extend infinitely transverse to the $z$ axis. The fluid extends to $+ \infty$ in the $z$ direction and the prism extends to $- \infty$ in  $z$ direction. }
\label{TKR}
\end{figure}

Suppose that a CSTF of thickness $\Ldiel$ is used as the partnering dielectric material in the TKR configuration and the planar metal/prism interface is identified as the  plane $z=0$. The CSTF comprises closely nested nanohelixes of a dielectric material that were grown parallel to each other by the process of physical vapor deposition. The CSTF is infiltrated by a fluid of refractive index $\ninf$ which also fills the half space $z>\Lmet+\Ldiel$. The prism material is taken to fill the half space $z<0$.
A schematic is provided in Fig. \ref{TKR}.

The anisotropy and nonhomogeneity of the CSTF are macroscopically quantified by the relative permittivity dyadic \cite{Mackay_2012,Lakh2006a}
\begin{equation}
\label{eqn:CSTF epsilon}
\=\eps_{\rm diel}(z)=  \=S_{\rm z}(z)\. \=S_{\rm y}(\chi)\. \epsoref 
\.\=S_{\rm y}^{-1}(\chi) \.\=S_{\rm z}^{-1}(z) \,.
\end{equation}
Here, the  local relative permittivity  dyadic 
\begin{equation}
\epsoref  =  \ux\ux \,\epsb  +  \uy\uy \,\epsc + \uz\uz \,\epsa 
\,
\end{equation}
in the material frame
captures the local orthorhombicity of the CSTF;
the  dyadic
\begin{equation}
\=S_{\rm z}(z)
=\uz\uz + \le \ux\ux+\uy\uy\ri\,\cos\le\frac{\pi z}{\Omega} \ri
+h\, \le \uy\ux-\ux\uy\ri\,\sin\le\frac{\pi z}{\Omega}\ri\,.
\end{equation}
 captures the rotation of  the local relative permittivity dyadic $\epsref=\=S_{\rm y}(\chi)\.\epsoref  \.\=S_{\rm y}^{-1}(\chi)$  in the laboratory frame about the $z$ axis, with $h\in\lec-1,1\ric$ denoting the structural handedness and $2\Omega$ the period; and the dyadic
\begin{equation}
 \=S_{\rm y}(\chi)=
 \uy\uy + (\ux\ux + \uz\uz) \, \cos\chi + (\uz\ux-\ux\uz)\,
\sin\chi
\end{equation}
represents the {\it locally\/} aciculate  
morphology of the CSTF,
with $\chi > 0$~deg being the rise angle  with respect to the $xy$ plane. Both $\epsref$ and $\epsoref$
have the same eigenvalues---denoted by $\epsa$, $\epsb$, and $\epsc$---but their eigenvectors differ. All three parameters   $\eps_{\rm a,b,c}$ of the fluid-infiltrated CSTF depend not only on $\lambdao$ but also  on $\ninf$
(which is itself dependent on $\lambdao$) \cite{Lakh2006a}. If $\eps_{\rm a,b,c}$ are known for $\ninf=1$
 at a specific value of $\lambdao$, then a combination of inverse and
 forward Bruggeman homogenization formalisms can be used to deduce their values for $\ninf\ne1$ at the same $\lambdao$ \cite{Mackay_2012}.

The electric field phasor of the plane wave incident on the prism/metal interface can be written as
\cite{Lakh2006a}
\begin{eqnarray}
\nonumber
\Einc&=& \left\{\as \left(-\ux\spsi + \uy \cpsi\right) +\ap
\left[-\left( \ux \cpsi + \uy \spsi \right) \ctheta  + \uz \stheta\right]
\right\}
\\[5pt]
&&\quad\times\exp\left[ i\ko\nprism \left( x\cpsi + y\spsi \right)\stheta\right]
 \,\exp\left(i\ko\nprism z\ctheta\right)\,,
\end{eqnarray}
where $i=\sqrt{-1}$, $\as$ is the amplitude of the $s$-polarized component and $\ap$ of the $p$-polarized component.
The incidence direction is specified by the angles $\theta\in\left[0^\circ,90^\circ\right)$ and
$\psi\in\left[0^\circ,360^\circ\right)$.
The electric field phasor of the reflected plane wave can be written as \cite{Lakh2006a}
\begin{eqnarray}
\nonumber
\Erefl&=& \left\{\rs \left(-\ux\spsi + \uy \cpsi\right) +\rp
\left[\left( \ux \cpsi + \uy \spsi \right) \ctheta  + \uz \stheta\right]
\right\}
\\[5pt]
&&\quad\times\exp\left[ i\ko\nprism \left( x\cpsi + y\spsi \right)\stheta\right]
 \,\exp\left(-i\ko\nprism z\ctheta\right)\,,
\end{eqnarray}
where $\rs$ is the amplitude of the $s$-polarized component and $\rp$ of
the $p$-polarized component. The reflectance is defined as
\begin{equation}
R=\frac{\vert\rs\vert^2+\vert\rp\vert^2}{\vert\as\vert^2+\vert\ap\vert^2}\,.
\label{Rdef}
\end{equation}

The procedure to compute $\rs$ and $\rp$, and therefore $R$,  from $\as$ and $\ap$ as functions
of $\lambdao$, $\theta$, and $\psi$ is available
in detail elsewhere \cite{PMLbook}. The procedure to obtain $\eps_{\rm a,b,c}$ as functions
of $\lambdao$ and $\ninf$ is also available in detail elsewhere \cite{Mackay_2012,Lakh2006a}.

\subsection{Artificial Neural Networks}
\label{sect:ANN}
 
ANNs are  machine-learning algorithms of a specific type \cite{Goodfellow_2016}. Machine-learning algorithms are constructed such that they improve in performance over time for a specific task without being explicitly programmed for that task. Typically, an ANN comprises multiple nodes (neurons) organized in several  layers arranged in a hierarchy. Each node in a given layer is interconnected with all the nodes in the adjacent layers. These interconnections are represented as numerical values called weights. The designated first layer  serves as the \textit{input layer} and the designated last layer  as the \textit{output layer}. 
Each node in a given layer computes the linear combination of the value of each node in the previous layer, along with that node's  weight. In order to account for possible non-linear processes, the linear combination is fed into an activation function, such as the sigmoid or rectifier function. \red{A neuron with no activation function is called a linear neuron, and an ANN composed exclusively of linear neurons will fit a linear model to the data.}

ANNs require training and testing before implementation. An ANN is trained on a data set $\mathbb{R}$ of  column vectors 
denoted by $ \boldsymbol{R}_j$, $ j \in [1,J] $. $\boldsymbol{R}_j$ consists of $K$ scalars $R_{jk}$, $ k \in [1,K] $. In addition,  there are label vectors $ \boldsymbol{n}_{j} $,   each consisting of $L$ scalars $n_{j\ell}$, $\ell \in [1,L] $. Every $ \boldsymbol{R}_j$ is accompanied by a unique 
$\boldsymbol{n}_{j}$, but some of the labels may share the same value when considering noisy data. As an example, two different sensor chips employed in the TKR configuration may produce slightly different reflectance signatures given the same infiltrating liquid, due to differences in sensor-chip quality from the fabrication process.

For a specific $j = j^\prime$, $\boldsymbol{R}_{j^\prime}$ is fed into the input layer, $ R_{j^\prime k} $
being fed to the node labeled $k$  in that layer. With the weights randomized, the label vector $ \boldsymbol{n}_{j^\prime}^\star $ is predicted by the ANN and an error value is calculated based on some predefined error function of $ \boldsymbol{n}_j $ and $ \boldsymbol{n}_j^\star $.  In general, $ \boldsymbol{n}_j^\star \ne \boldsymbol{n}_{j^\prime}^\star$.
 This error function can be expressed as a function of the weights, since  $ \boldsymbol{n}_j^\star $ is a function of the weights. Typically, an ANN  uses a gradient-descent method  \red{\cite{Rumelhart_1987}} to find the   weights such that the average error over $\mathbb{R}$ is sufficiently small. 
The ANN is then tested on a data set $\overline{\mathbb{R}}$. 

In this paper, every  testing vector and its elements are identified by the addition of an overbar to the symbol for the corresponding training vector and its elements. In addition, we use $\mathbb{R}^{a}_{b}$ with $a$ and $b$ as placeholders to denote the polarization state (represented by the ratio of $\as$ to $\ap$) and the angle $\psi$, respectively, of the incident plane wave for which the reflectance data
are obtained. This notation is explained in Tables \ref{table: polstate} and  \ref{table: psi}.  Next,
all styles and fonts of uppercase `R'  represent reflectance data calculated for various polarization states, angles $\theta$, and angles $\psi$. Finally, all styles and fonts of lowercase `n'  represent refractive-index data (i.e., $\ninf$) corresponding to the reflectance data. Let us note that $\ninf$ denotes the refractive index of the infiltrating liquid in general, whether used for training, testing, both, or neither.

		\begin {table}[h]
		\caption {\label{table: polstate}  Label $a$ and polarization state of the incident plane wave in the prism.} 
		\begin{center}
			\begin{tabular}{p{2.5cm} p{1.6cm} p{1.6cm}p{5cm}}
				\hline
				\hline
				$\rm  $ $a$ &  $\as$  & $\ap$  & $\rm polarization \ state$  \\
				\hline
					
				1 & $0$ & $1$ & linear ($p$)\\
		
					2  & $1$ & $0$ & linear ($s$) \\
			
					3 & $1/\sqrt{2}$ & $1/\sqrt{2} $ & linear (a mixture of $p$ and $s$)  \\
										
					4 & $1/\sqrt{2}$ & $-1/\sqrt{2}$ & linear (another mixture of $p$ and $s$)  \\
				 5 & $1/\sqrt{2}$ & ${i}/\sqrt{2}$ & left circular  \\
										
					6 & $1/\sqrt{2}$ & $- {i}/\sqrt{2}$ & right circular  \\
				
				\hline
				\hline
			\end{tabular}
		\end{center}
		\end {table}
		
			\begin {table}[h]
			\caption {\label{table: psi} 
			Label $b$ and angle $\psi$ chosen for the incident plane wave in the prism.} 
			\begin{center}
				\begin{tabular}{p{2.5cm} p{1.6cm} p{1.6cm}p{4cm}}
					\hline
					\hline
					$\rm  $ $b$  &  $\psi$~(deg)  \\
					\hline
						
					1 &  $0$\\
			
						2  & $18$ \\
				
						3 & $36$\\
											
						4 & $54$\\
					 5 & $72$\\
											
						6 & $90$\\
					
					\hline
					\hline
				\end{tabular}
			\end{center}
			\end {table}


Theorems of machine learning suggest that an algorithm tailored to the needs of the specific application must be sought \cite{Wolpert_1996,Wolpert_1997}.
Therefore, we trained two ANNs with differing structures for various $ \mathbb{R} $ with label vectors $ \boldsymbol{n}_{j} $ and tested the ANNs for various $ \overline{\mathbb{R}} $  with the label vectors $ \overline{\boldsymbol{n}}_{j} $. For this work, both $ \boldsymbol{n}_{j} $ and $ \overline{\boldsymbol{n}}_{j} $ comprise just one element each (i.e., $L=1$) and therefore are denoted  simply as $ n_{j} $ and $ \overline{n}_{j} $, respectively.


\section{Simulations}
\label{sect:Exp}


 \subsection{Training Data}
 \label{sect: training data}
 
After fixing $\lambdao = 635 $~nm,
various $ \mathbb{R} $ were calculated using a CSTF with half-period $\Omega = 200$~nm and thickness $\Ldiel = 1200$~nm made of titanium oxide. 
When $\ninf=1$,
we set $\epsa=2.1395$, $\epsb=3.6690$, and $\epsc=2.8257$, corresponding to  $ \chi = 37.6745^\circ$,
as provided by actual measurements on columnar thin films \cite{HWH}.
Values of $\eps_{\rm a,b,c}$ for $\ninf>1$ were computed using a homogenization approach detailed
elsewhere \cite{Mackay_2012}. Choosing the metal to be silver, we fixed $\nmet = \sqrt{\epsmet}= 0.18016 +  i3.4531$ \cite{Au_index}
and $\Lmet=30$~nm \cite{Swiontek_2013}. The prism was chosen to be made of SF11 glass so that
$\nprism = 1.78471$.
 Noting that most applications for biosensors involve small concentrations of analytes dispersed in solvents of refractive index close to that of water ($\nwater=1.333$), we chose to calculate various $\mathbb{R}$ and $ \overline{\mathbb{R}} $ for $\ninf \in [1.3,1.4]$ for equal increments $\Delta\ninf = 0.0\overline{111}$ and $\Delta\ninf \approx 0.000503$, respectively, as determined by $J$.  All 
 $\mathbb{R}$ and $ \overline{\mathbb{R}} $ were calculated for $\phi \in [ -25^\circ, 24.8^\circ ] $ in steps of $ \Delta \phi = 0.2^\circ $,  which can be related  \cite{Swiontek_2016}  to $\theta$ by the standard law of refraction, yielding $ \theta = 45^\circ + \sin^{-1}\left(\nprism^{-1} \sin\phi\right) $
for the prism with a $45^\circ$--$90^\circ$--$45^\circ$ triangle as its cross section. \red{All data were computed using Mathematica\textsuperscript{\textregistered} running on a  laptop computer with 
a 4-core 2.4-GHz processor and 16~GB of RAM. With $\phi$, $\ninf$, and $b$ fixed,
the average estimated computation time was $\sim 0.06$~s for the set of  six polarization states specified in
 Table~\ref{table: polstate}.}

\subsubsection{Training with Incident Light of  Diverse Polarization States}

SPP-wave modes are capable of being excited by incident light of an arbitrary polarization state in the prism-coupled configuration when a periodically nonhomogeneous dielectric material is partnering a metal thin film \cite{PMLbook}. However, experiments have shown that using $p$-polarized incident light results in a higher sensitivity over $s$-polarized incident light, based on the definition \cite{Swiontek_2016}
\begin{equation}
\rho = \frac{ \thetaSPP(n_{\rm inf_2}) - \thetaSPP(n_{\rm inf_1}) }{ n_{\rm inf_2} - n_{\rm inf_1} } 
\label{eqn: sensitivity}
\end{equation}
of sensitivity,
 where $ n_{\rm inf_\ell} $, $\ell \in \{1,2\} $, is the refractive index of an infiltrant fluid labeled $\ell$ and $\thetaSPP(n_{\rm inf_\ell})$ is the value of $\thetaSPP$ for $n_{\rm inf_\ell}$. One might then infer that $\mathbb{R}$ from $p$-polarized incident light will yield better training for an ANN. However, since $\rho$ is focused on SPP-wave modes but not on waveguide modes and other phenomena, as  mentioned in Sec.~\ref{sect:Intro}, we conjectured that $s$-polarized incident light as well as incident light of other polarization states may also be useful for ANN training.
 
Therefore, our first group of simulations used  $ \mathbb{R} $ calculated for six different  polarization states of incident light with $\psi = 0^\circ$. Consistently with Tables~\ref{table: polstate} and ~\ref{table: psi}, the six training data sets are
denoted by $\mathbb{R}_{1}^{a} $, $a\in[1,6]$.
For these data sets, we fixed $J = 10$ and $ K=250$; thus, $10$ values of  $n_j$ and $250$ values of $\theta$ were used. For  $ \overline{\mathbb{R}}_{1}^{a} $, $a\in[1,6]$, we set  $ J = 200 $ and $K = 250$.
 
 \subsubsection{Training with  Light Incident at Diverse $\psi$ }
 
 To determine what value of the angle $\psi$   is the most effective,  training data sets 
 $\mathbb{R}^{1}_{b}$,  $b\in[1,6]$,
were implemented for separate ANNs. For these sets, we fixed $J = 10$ and $ K=250$. For the counterpart testing data sets $ \overline{\mathbb{R}}^{1}_{b} $, $b\in[1,6]$, we used $ J = 200 $ and $K=250$.
 
\subsubsection{Training with Some  Combinations of Diverse Polarization States}
\label{sect: pol comb}

Even if one polarization state is more effective than the others, the ANN still may benefit from the inclusion of reflectance data from the other polarization states. We use the notation $ \mathbb{R}^{1:c}_{b} $ to mean reflectance data that includes polarization states labeled
 $a\in\left\{1,2,....,c\right\}$; thus, $ \mathbb{R}^{1:c}_{b} =  \mathbb{R}^{1}_{b} \cup \mathbb{R}^{2}_{b} \cup\dots\cup\mathbb{R}^{c}_{b}
 $ and
  $ \mathbb{R}^{1:1}_{1} \equiv \mathbb{R}^{1}_{1} $.
 Fixing $\psi=0^\circ$, we focused on $ \mathbb{R}^{1:c}_{1} $. For $ \mathbb{R}^{1:1}_{1} $, $ \mathbb{R}^{1:2}_{1} $, $ \mathbb{R}^{1:3}_{1} $, $ \mathbb{R}^{1:4}_{1} $, $ \mathbb{R}^{1:5}_{1} $, and $ \mathbb{R}^{1:6}_{1} $, we have $J = 10$ and $ K = 250,\, 500,\, 750, \, 1000,\, 1250$, 
 and $1500$, respectively. Whereas  $ J = 200 $ for $ \overline{\mathbb{R}}^{1:c}_{1}$,
 $ K = 250,\, 500,\, 750, \, 1000,\, 1250$, 
 and $1500$ for $c = 1,\,2,\,3,\,4,\,5$, and $6$, respectively.

\subsubsection{Training with Some  Combinations of Diverse $\psi$}

With similar reasoning as in Sec.~\ref{sect: pol comb}, we write $ \mathbb{R}^{a}_{1:c} $ to mean reflectance data that includes 
values of $\psi$ corresponding to $b\in\left\{1,2,....,c\right\}$; thus, $ \mathbb{R}^{a}_{1:c} =  \mathbb{R}^{a}_{1} \cup \mathbb{R}^{a}_{2} \cup\dots\cup\mathbb{R}^{a}_{c}
 $ and
$ \mathbb{R}^{1}_{1:1} \equiv \mathbb{R}^{1}_{1} \equiv \mathbb{R}^{1:1}_{1} $. 
Fixing our attention only on incident   $p$-polarized light, we focused on $ \mathbb{R}^{1}_{1:c} $. For $ \mathbb{R}^{1}_{1:1} $, $ \mathbb{R}^{1}_{1:2} $, $ \mathbb{R}^{1}_{1:3} $, $ \mathbb{R}^{1}_{1:4} $, $ \mathbb{R}^{1}_{1:5} $, and $ \mathbb{R}^{1}_{1:6} $, $J = 10$ and and $ K = 250,\, 500,\, 750, \, 1000,\, 1250$, 
 and $1500$, respectively. The testing data set $ \overline{\mathbb{R}}^{1}_{1:c} $ was constituted with $ J = 200 $ and $ K = 250,\, 500,\, 750, \, 1000,\, 1250$, 
 and $1500$ for $c = 1,\,2,\,3,\,4,\,5$, and $6$, respectively.

\subsubsection{Training with Combinations of Polarization State   and $ \psi $ }

Of the multitude of reflectance data sets that can be determined for combinations of polarization states \textit{and}  $\psi$, we only investigated   \begin{equation}
\mathbb{R}^{1:6}_{1:6} =
\left(\mathbb{R}^{1}_{1} \cup \mathbb{R}^{1}_{2} \cup\dots\cup\mathbb{R}^{1}_{6}\right)\cup
\left(\mathbb{R}^{2}_{1} \cup \mathbb{R}^{2}_{2} \cup\dots\cup\mathbb{R}^{2}_{6}\right)\cup\dots\cup
\left(\mathbb{R}^{6}_{1} \cup \mathbb{R}^{6}_{2} \cup\dots\cup\mathbb{R}^{6}_{6}\right)\,.
\end{equation}
For this training data set, $ J = 10 $ and $ K = 9000 $. For the 
corresponding testing data set $ \overline{\mathbb{R}}^{1:6}_{1:6}$, $J = 200$ and $ K = 9000 $.

\subsubsection{Addition of Noise}

\begin{figure}
\centering
\includegraphics[scale=0.6]{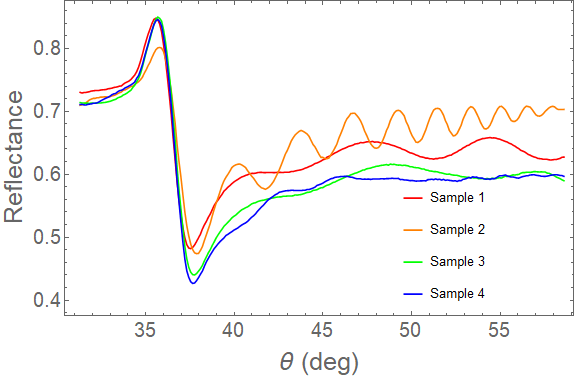}
\caption{Reflectance as a function of $\theta$ measured for four different samples in the TKR configuration, each with a $30$-nm-thick aluminum film and air as the partnering dielectric material.
Each data point from a given plot was compared to the  data point for the same $\theta$ from all other plots. The absolute difference of all these measurements was averaged and calculated to be $0.0409$. For good measure, this was rounded up to $0.05$. This magnitude was split about $0$, yielding $-0.025$ for the lower  bound and $0.025$ for the upper  bound of random numbers used to add noise to $\mathbb{R}^1_6$. }
\label{SMPR}
\end{figure}

For the data set that yielded the best ANN performance, noise was added to that set to simulate experimental data. The corresponding noisy data set is denoted by $ \mathds{R} $ with the specific superscript and subscript of the set that yielded the best performance. As shown in Sec.~\ref{sect:Results},
the best-performance set was $\mathbb{R}^1_6$. The noise was simulated as a random number between $ -0.025$ and $0.025 $ added to each element $ R_{jk} \in \boldsymbol{R}_j \in \mathbb{R}^1_6 $ for $10$ different instances. The upper and lower bounds of the random numbers were chosen based on the average absolute difference of reflectance data between four different $30$-nm-thick aluminum films implemented in a specific TKR apparatus, as shown in Fig. \ref{SMPR} . Thus, for $ \mathds{R}^1_6 $, $ J = 10 \times 10 = 100$ and $ K = 250$. For the corresponding testing data set $ \overline{\mathds{R}}^1_6$, $ J = 200 \times 10 = 2000$ and $ K = 250$. 

\subsubsection{Training with just SPR data}

Up until this stage, each training and testing data set has included reflectance data from $\theta$ ranging from
about $30^\circ$ to about $60^\circ$. With inspiration from conventional SPR occurring at the interface of a metal and a homogeneous dielectric material, we devised a data set  denoted by $ \mathbb{R}^{SPR} $ with $\theta$ ranging between $0^\circ$ and $70^\circ$ in steps of $0.1^\circ$, with $a=1$, $b=1$, $J = 10$, $K = 701$, and with no noise added. In order to ensure that only the occurrences of SPR were captured in $ \mathbb{R}^{SPR}$, the reflectances were calculated with $\Ldiel = 1200$~nm and $\Ldiel = 1600$~nm, and only those dips that differed by less than $0.5^\circ$ for the two values of $\Ldiel$  were retained \cite{Swiontek_2013,Swiontek_2016}. Thus most entries in the column vectors in $ \mathbb{R}^{SPR} $ were zero.
The corresponding testing data set $ \overline{\mathbb{R}}^{SPR} $ was constituted with $ J = 200 $, and $ K = 701$. This data set would allow us to to determine the relative efficacy of SPR alone for training ANNs.

\subsection{ANN  Parameters}
\label{sect:Net Param}

\red{All training and testing was done using MATLAB\textsuperscript{\textregistered} running on a laptop computer with 
a 4-core 2.4-GHz processor and 16~GB of RAM. In all, training took no more than a day.} Each training data set was used for two separate ANN structures. The first type of ANN structure, denoted by ANN$_1$, consisted of three layers: an input layer, one hidden layer, and an output layer. The hidden layer contained 100 nodes with no \red{activation function (linear neurons)} and the output layer contained one node. The second ANN structure, denoted by ANN$_2$ consisted of four layers: an input layer, two hidden layers, and an output layer. The two hidden layers each contained $100$ nodes with \red{rectifier activation functions}, and the output layer contained one node. The input layer for both ANN$_1$ and ANN$_2$ consisted of one vector. The size of this vector was either $250$, $500$, $750$, $1000$, $1250$, $1500$, $9000$, or $701$ depending on the specific data set being used for training. The \red{stochastic gradient-descent with momentum method \cite{Patterson_2017} was chosen for optimization with an initial learning rate of $0.01$ and the error function defined as the mean-squared error. The number of maximum epochs was $10,000$, an epoch being the number of times the learning algorithm was exposed to a given training data set.} 
 
The mean $\mu$,  standard deviation $\delta$, median $\sigma $, maximum $M$, and minimum $m$ of $ \lvert \overline{n}_{j}^\star - \overline{n}_j \rvert$ for each testing data set were used to assess the performance of each ANN. In addition, for several instances of training for any given ANN with a particular structure and training data set, the performance of that ANN given the same testing data set may vary due to the fact that the weights are randomized at the start of each training instance. Therefore, given a particular training data set, we trained each ANN for ten instances and averaged the aforementioned statistical measures for all of the training instances. Thus, hereafter, the symbols $\mu$, $\delta$, $\sigma$, $M$, and $m$,  denote  performance measures averaged over ten training instances. Ideally, all five 
performance measures should be as close to $0$ as possible.


\section{Numerical Results and Discussion}
\label{sect:Results}

\subsection{ANN$_1$}

Values of all five performance measures for every training data set for ANN$_1$ are listed in Table \ref{table: measures}. 
This table is divided into six blocks, one each for  $ \mathbb{R}^{a}_{1} $, $ \mathbb{R}^{1}_{b} $, $ \mathbb{R}^{1:c}_{1} $, $ \mathbb{R}^{1}_{1:c} $, $ \mathbb{R}^{1:6}_{1:6} $, and $ \mathds{R}^{1}_{6} $ ($1,2,3,4,5,6$, respectively). Within each of the first four blocks, those training sets which yielded the lowest value for each performance measure are highlighted by a colored background. In addition, the overall lowest value for each performance measure is identified in boldface.

\begin{table}[h]
\centering
\caption{Values for $\mu$, $\delta$, $\sigma$, $M$, and $m$ for ANN$_1$. Note that $ \mathbb{R}^{1}_{1}$, $ \mathbb{R}^{1:1}_{1}$, and $ \mathbb{R}^{1}_{1:1}$ are equivalent by virtue of the definitions in Sec. \ref{sect: training data}.}\vspace{2mm}
\label{table: measures}
\resizebox{\columnwidth}{!}{ %
\begin{tabular}{|c|c|c|c|c|c||||c|c|c|c|c|c|c|}
\hline
 & $\mu$ ($10^{-4}$) & $\delta$ ($10^{-4}$) & $\sigma$ ($10^{-4}$) & $M$ ($10^{-4}$) & $m$ ($10^{-4}$) &  & $\mu$ ($10^{-4}$) & $\delta$ ($10^{-4}$) & $\sigma$ ($10^{-4}$) & $M$ ($10^{-4}$) & $m$ ($10^{-4}$)\\
  \hline
$ \mathbb{R}^{1}_{1}$& \cellcolor{red}$5.0071$ & \cellcolor{red}$5.0024$ & \cellcolor{red}$3.4032$ & \cellcolor{red}$21.6967$ & \cellcolor{red}$0.0224$ &  $ \mathbb{R}^{1}_{1}$ & $5.0071$ & $5.0024$ & $3.4032$ & $21.6967$ & $0.0224$\\
\hline
$ \mathbb{R}^{2}_{1}$ & $11.3230$ & $10.4618$ & $8.2767$ & $46.2595$ & $0.0689$ & $ \mathbb{R}^{1}_{2}$  & $5.7570$ & $5.0929$ & $4.2061$ & $21.3344$ & \cellcolor{orange}\textbf{0.0174} \\
\hline
$ \mathbb{R}^{3}_{1}$ & $12.5241$ & $11.6265$ & $8.9733$ & $49.7426$ & $0.0719$ & $ \mathbb{R}^{1}_{3}$  & $5.2991$ & $5.2920$ & $3.7781$ & $22.5849$ & $0.0294$ \\
\hline
$ \mathbb{R}^{4}_{1}$ & $7.6926$ & $6.3344$ & $6.1033$ & $26.3900$ & $0.0698$ & $ \mathbb{R}^{1}_{4}$  & $5.4146$ & $5.5531$ & \cellcolor{orange}$3.2702$ & $22.2373$ & $ 0.0290 $ \\
\hline
$ \mathbb{R}^{5}_{1}$ & $ 7.8289 $ & $6.7660$ & $5.8249$ & $29.2593$ & $ 0.0456 $ & $ \mathbb{R}^{1}_{5}$  & $5.0476$ & $4.7477$ & $3.6241$ & $19.8604$ & $0.0207$ \\
\hline
$ \mathbb{R}^{6}_{1}$ & $11.1633$ & $10.1215$ & $8.1131$ & $43.5510$ & $0.0528$ & $ \mathbb{R}^{1}_{6}$  & \cellcolor{orange}\textbf{4.9258} & \cellcolor{orange}\textbf{4.6006} & $3.5026$ & \cellcolor{orange}\textbf{19.7462} & $ 0.0392 $\\
 \hline
 \hline
 \hline
$ \mathbb{R}^{1:1}_{1}$ & \cellcolor{yellow}$5.0071$ & \cellcolor{yellow}$5.0024$ & \cellcolor{yellow}$3.4032$ & \cellcolor{yellow}$21.6967$ & \cellcolor{yellow}$0.0224$ & $ \mathbb{R}^{1}_{1:1}$  & \cellcolor{green}$5.0071$ & \cellcolor{green}$5.0024$ & $3.4032$ & \cellcolor{green}$21.6967$ & $0.0224$ \\
\hline
$ \mathbb{R}^{1:2}_{1}$ & $8.0585$ & $7.3837$ & $5.7144$ & $31.6771$ & $0.0399$ & $ \mathbb{R}^{1}_{1:2}$  & $5.3233$ & $5.3843$ & $3.4034$ & $22.4634$ & \cellcolor{green}$0.0175$\\
\hline
$ \mathbb{R}^{1:3}_{1}$ & $ 10.0389 $ & $ 8.8401 $ & $ 7.8057 $ & $ 40.6256 $ & $ 0.0378 $ & $ \mathbb{R}^{1}_{1:3}$  & $5.7015$ & $6.3600$ & \cellcolor{green}\textbf{3.1690} & $26.0298$ & $0.0218$ \\
\hline
$ \mathbb{R}^{1:4}_{1}$ & $10.6654$ & $9.7212$ & $8.0245$ & $ 42.1574 $ & $ 0.0486 $ & $ \mathbb{R}^{1}_{1:4}$  & $6.2201$ & $6.0222$ & $4.5226$ & $ 25.1908 $ & $ 0.0392 $\\
\hline
$ \mathbb{R}^{1:5}_{1}$ & $10.3369$ & $9.3632$ & $7.5046$ & $39.6058$ & $0.0324$ & $ \mathbb{R}^{1}_{1:5}$  & $6.3917$ & $ 5.9051 $ & $4.6640$ & $ 22.9499 $ & $ 0.0223 $\\
\hline
$ \mathbb{R}^{1:6}_{1}$ & $11.6914$ & $10.2831$ & $8.7284$ & $46.9326$ & $0.0498$ & $ \mathbb{R}^{1}_{1:6}$  & $6.6261$ & $6.0738$ & $ 4.9742 $ & $ 25.2392 $ & $ 0.0281 $\\
\hline
\hline
\hline
$ \mathbb{R}^{1:6}_{1:6}$ & $19.4394$ & $16.9059$ & $14.9430$ & $67.7048$ & $0.0498$ & $ \mathds{R}^{1}_{6}$  & $ 8.3261 $  & $ 6.6368 $ & $ 6.7558 $ & $36.9478$ & $ 0.0051 $ \\
\hline
\end{tabular}}
\end{table}

\subsection{ANN$_2$}

Values of all five performance measures for every training data set for ANN$_2$ are listed in Table \ref{table: measures2}. 
This table is divided into six blocks, one each for  $ \mathbb{R}^{a}_{1} $, $ \mathbb{R}^{1}_{b} $, $ \mathbb{R}^{1:c}_{1} $, $ \mathbb{R}^{1}_{1:c} $, $ \mathbb{R}^{1:6}_{1:6} $, and $ \mathds{R}^{1}_{6} $ ($1,2,3,4,5,6$, respectively). Within each of the first four blocks, those training sets which yielded the lowest value for each performance measure are highlighted by a colored background. In addition, the overall lowest value for each performance measure is identified in boldface.

\begin{table}[h]
\centering
\caption{Values for $\mu$, $\delta$, $\sigma$, $M$, and $m$ for ANN$_2$. Note that $ \mathbb{R}^{1}_{1}$, $ \mathbb{R}^{1:1}_{1}$, and $ \mathbb{R}^{1}_{1:1}$ are equivalent by virtue of the definitions in Sec. \ref{sect: training data}.}\vspace{2mm}
\label{table: measures2}
\resizebox{\columnwidth}{!}{ %
\begin{tabular}{|c|c|c|c|c|c||||c|c|c|c|c|c|c|}
\hline
 & $\mu$ ($10^{-4}$) & $\delta$ ($10^{-4}$) & $\sigma$ ($10^{-4}$) & $M$ ($10^{-4}$) & $m$ ($10^{-4}$) &  & $\mu$ ($10^{-4}$) & $\delta$ ($10^{-4}$) & $\sigma$ ($10^{-4}$) & $M$ ($10^{-4}$) & $m$ ($10^{-4}$)\\
  \hline
$ \mathbb{R}^{1}_{1}$& $18.4474$ & \cellcolor{red}$13.0167$ & $15.7931$ & $73.7659$ & $0.1729$ &  $ \mathbb{R}^{1}_{1}$ & \cellcolor{orange}$18.4474$ & $13.0167$ & \cellcolor{orange}$15.7931$ & $73.7659$ & $0.1729$\\
\hline
$ \mathbb{R}^{2}_{1}$ & \cellcolor{red}$18.2933$ & $15.2521$ & \cellcolor{red}$14.7095$ & \cellcolor{red}$63.1350$ & \cellcolor{red}$0.0731$ & $ \mathbb{R}^{1}_{2}$  & $19.3895$ & $13.8907$ & $16.9469$ & $76.9789$ & \cellcolor{orange}$0.1208$ \\
\hline
$ \mathbb{R}^{3}_{1}$ & $151.3610$ & $90.8141$ & $145.3446$ & $337.8720$ & $0.8946$ & $ \mathbb{R}^{1}_{3}$  & $18.8519$ & $13.1368$ & $17.0860$ & $74.3209$ & $0.1912$ \\
\hline
$ \mathbb{R}^{4}_{1}$ & $180.7079$ & $112.1547$ & $172.4971$ & $393.8738$ & $0.5904$ & $ \mathbb{R}^{1}_{4}$  & $20.6076$ & $15.1996$ & $18.1812$ & $82.8628$ & $ 0.1438 $ \\
\hline
$ \mathbb{R}^{5}_{1}$ & $ 196.3213 $ & $ 119.1359 $ & $ 190.0017 $ & $415.9390$ & $ 0.9442 $ & $ \mathbb{R}^{1}_{5}$  & $ 20.4725 $ & $14.5558$ & $17.6106$ & $80.5762$ & $0.2359$ \\
\hline
$ \mathbb{R}^{6}_{1}$ & $185.2945$ & $112.0894$ & $178.7833$ & $393.0062$ & $0.7333$ & $ \mathbb{R}^{1}_{6}$  & $17.3366$ & \cellcolor{orange}$11.1896$ & $16.6813$ & \cellcolor{orange}$65.9253$ & $ 0.2087 $\\
 \hline
 \hline
 \hline
$ \mathbb{R}^{1:1}_{1}$ & $18.4474$ & $13.0167$ & $15.7931$ & $73.7659$ & $0.1729$ & $ \mathbb{R}^{1}_{1:1}$ & $18.4474$ & $13.0167$ & $15.7931$ & $73.7659$ & $0.1729$ \\
\hline
$ \mathbb{R}^{1:2}_{1}$ & $8.3502$ & $6.5265$ & $6.9467$ & $33.3071$ & \cellcolor{yellow}$0.0407$ & $ \mathbb{R}^{1}_{1:2}$  & $8.6354$ & $7.1220$ & $6.8011$ & $44.4590$ & $0.0571$\\
\hline
$ \mathbb{R}^{1:3}_{1}$ & $ 7.8625 $ & $ 6.2680 $ & \cellcolor{yellow}$ 6.2770 $ & $ 29.3114 $ & $ 0.0551 $ & $ \mathbb{R}^{1}_{1:3}$  & $7.3186$ & $5.6583$ & $6.1465$ & $32.9818$ & $0.0510$ \\
\hline
$ \mathbb{R}^{1:4}_{1}$ & $8.0364$ & $6.3230$ & $6.5755$ & $ 27.4354 $ & $ 0.1191 $ & $ \mathbb{R}^{1}_{1:4}$  & $6.8830$ & $5.3598$ & $5.6107$ & $ 27.5089 $ & $ 0.0497 $\\
\hline
$ \mathbb{R}^{1:5}_{1}$ & \cellcolor{yellow}$7.7478$ & \cellcolor{yellow}$5.9803$ & $6.2856$ & \cellcolor{yellow}$24.7017$ & $0.0493$ & $ \mathbb{R}^{1}_{1:5}$  & $6.8034$ & \cellcolor{green}\textbf{4.9553}  & $5.6408$ & $ 24.6009 $ & $ 0.0552 $\\
\hline
$ \mathbb{R}^{1:6}_{1}$ & $8.3639$ & $6.3646$ & $7.0208$ & $26.1711$ & $0.0625$ & $ \mathbb{R}^{1}_{1:6}$  & \cellcolor{green}\textbf{6.5693} & $5.2932$ & \cellcolor{green}\textbf{5.3522} & \cellcolor{green}\textbf{23.8877} & \cellcolor{green}$ 0.0250 $\\
\hline
\hline
\hline
$ \mathbb{R}^{1:6}_{1:6}$ & $9.9815$ & $9.2920$ & $6.9350$ & $39.3160$ & $0.0527$ & $ \mathds{R}^{1}_{6}$  & $17.6296 $  & $ 12.4912 $ & $ 15.8691 $ & $ 78.4342 $ & $0.0130$ \\
\hline
\end{tabular}}
\end{table}

\subsection{Comparison of ANN$_1$ and ANN$_2$}

Training of ANN$_1$ for each $\mathbb{R}$ provided a linear function to predict $\ninf$. By having three out of five performance measures as the lowest, we may define $\mathbb{R}^{1}_{6}$ as providing the best training for ANN$_1$. We also note that changing $\psi$ does not change the performance of ANN$_1$ as much as changing the polarization state. The same statement goes for the addition of more $\psi$ values versus the addition of more polarization states.

Training of ANN$_2$ for each $\mathbb{R}$ provided a non-linear function to predict $\ninf$. By having $4$ out of $5$ performance measures being the lowest, we may define $\mathbb{R}^{1}_{1:6}$ as providing the best training for ANN$_2$. If we compare ANN$_2$ and $\mathbb{R}^{1}_{1:6}$ with ANN$_1$ and $\mathbb{R}^{1}_{6}$, the latter pair perform better, despite the former training set containing more data and the latter ANN having a simpler structure.

Training of ANN$_1$ with $ \mathbb{R}^{SPR} $ yielded $\mu = 220.1760 \times 10^{-4} $ , $\delta = 156.1408 \times 10^{-4}$, $\sigma = 211.2664 \times 10^{-4}$, $m = 0.5761 \times 10^{-4}$, and $M = 507.8399 \times 10^{-4}$, while training of ANN$_2$ with $ \mathbb{R}^{SPR} $ yielded $\mu = 251.2219 \times 10^{-4} $ , $\delta = 145.3823 \times 10^{-4}$, $\sigma = 251.2638 \times 10^{-4}$, $m = 2.3957 \times 10^{-4}$, and $M = 499.3997 \times 10^{-4}$. Overall, both sets of measures are significantly worse compared to training with any other $ \mathbb{R} $. We conclude from this that sensing only with SPR data seriously undermines the ability of an ANN to correctly predict $\ninf$. 

When noise  consistent with   actual experimental data was added, we found that the performance of a   simple ANN with a small number of training examples  yielded no more than a $0.0037$ difference between the actual and predicted $ \ninf $ based on the fact that $M = 36.9478 \times 10^{-4}$ for ANN$_1$ trained with $ \mathds{R}^1_6$. Thus, use of ANNs provides a level of immunity to measurement noise.


\section{Concluding Remarks}
\label{sect:Conclusion} 

A typical method for sensing is via SPR using the TKR configuration. We have simulated reflectance data from a liquid-infiltrated CSTF partnering a metal thin-film in the TKR configuration and used this data to train two ANNs with differing structures. The performance measures of the ANNs for many training instance were compared. The various training data sets contained reflectance data calculated for various combinations of the polarization state
and the angle $\psi$. Some of this training data was complicated by realistic noise.  \red{We stress here that our work pertains directly to the sensing of $\ninf$  and only indirectly to the identification of the infiltrant fluid, functionalization \cite{Homola2008} being required
for the latter purpose.}

One main conclusion we have shown is that $\ninf$ can best be predicted from $p$-polarized light with $\psi = 90^\circ$, with an ANN having no activation function. This instance represents a best-case scenario. It will require the use of either a triangular prism with a very broad base or a hemispherical prism because the value of $\psi$ is very high.

Another main conclusion of this paper is that the inclusion of other reflectance data in addition to SPR data greatly improves the performance of an ANN. Given the simplicity and heuristic choice of the ANN$_1$ structure and the relative small number of training examples compared to the testing examples used for this work, we are optimistic that significant improvement in the performance can be achieved in the future by adding more training examples and refining the ANN$_1$ structure. Eventually, the application of ANNs may engender an era of simultaneous multianalyte sensing \cite{Swiontek_2013}. \red{We also expect our ANN methodology to apply when SPP waves
are manipulated by active functional materials and phase-change materials for enhanced sensitivity \cite{Rodrigo_2015,Sreek_2019}}.

\section*{\red{Appendix~A: MATLAB\textsuperscript{\textregistered} codes for Artificial Neural Networks}}
The variables \texttt{n} and \texttt{mb} represent the input layer size and mini-batch size, respectively. The mini-batch size was that of the training data set used for that instance.

\subsection*{A.1: ANN$_1$}
\begin{verbatim}
layers = [ ...
    sequenceInputLayer(n)
    fullyConnectedLayer(100)
    fullyConnectedLayer(1)
    regressionLayer];
options = trainingOptions('sgdm','InitialLearnRate',0.01, ...
    'MaxEpochs',10000,...
    'MiniBatchSize',mb)
    \end{verbatim}
\subsection*{A.2: ANN$_2$}
    \begin{verbatim}
    layers = [ ...
        sequenceInputLayer(n)
        fullyConnectedLayer(100)
        reluLayer
        fullyConnectedLayer(100)
        reluLayer
        fullyConnectedLayer(1)
        regressionLayer];
    options = trainingOptions('sgdm','InitialLearnRate',0.01, ...
        'MaxEpochs',10000,...
        'MiniBatchSize',mb)
\end{verbatim}

\noindent\textbf{Acknowledgments.}   
The research of  P.~D. McAtee and A. Lakhtakia is funded by the Charles Godfrey Binder Endowment at the Pennsylvania State University.


\end{document}